\documentclass{ws-procs9x6}

\newcommand{\gapr}{\raisebox{-.6ex}{\mbox{
$\stackrel{>}{\mbox{\scriptsize$\sim$}}\:$}}}
\newcommand{\lapr}{\raisebox{-.6ex}{\mbox{
$\stackrel{<}{\mbox{\scriptsize$\sim$}}\:$}}}
\def\ros{{\sl ROSAT}}
\def\asc{{\sl ASCA}}
\def\cha{{\sl Chandra}}
\begin{document}

\title{Thermal Radiation from Cooling Neutron 
Stars}

\author{G.~G. Pavlov}

\address{Pennsylvania State University,\\
525 Davey Lab,  
University Park, PA 16802, USA\\ 
E-mail: pavlov@astro.psu.edu}

\author{V.~E. Zavlin}
\address{Max--Planck--Institut f\"ur extraterrestrische Physik, \\
 Giessenbachstr.~1, 85748 Garching, Germany \\
E-mail: zavlin@mpe.mpg.de}

\maketitle

\abstracts{Observations of thermal radiation from neutron stars
 allow one to measure the surface temperatures
and confront them with cooling scenarios.
Detection of gravitationally redshifted spectral lines can yield
the mass-to-radius ratio.
In the few cases when the distance is known, one can measure
the neutron star radius, which is particularly important
to constrain the equation of state of the superdense matter
in the neutron star interiors. Finally, one can infer
the chemical composition of
the neutron star surface layers, which provides information about
formation of neutron stars and their interaction with the environments.
We present the observational results on thermal radiation
from active pulsars and
radio-quiet neutron stars, with emphasis on the results obtained with
the {\sl Chandra} X-ray Observatory
and discuss some implications of these results.
}

\section{Introduction}
%%%%%%%%%%%%%%%
Observations of thermal emission from 
the surface layers of isolated (non-accreting) neutron stars (NSs)
is among most important tools to study the nature of these exotic objects. 
However, only a small fraction of the currently observable NSs
allows direct observations of their surfaces.
First of all, these are middle-aged ($\tau\sim 10^4$--$10^6$ yr) 
radio and/or gamma-ray pulsars,
whose thermal radiation, with temperatures 0.3--1 MK, 
can dominate at soft X-ray and UV energies.
In addition to active pulsars, a number of radio-quiet
isolated NSs emitting thermal-like X-rays have been detected,
with typical temperatures $\sim 0.5$--5 MK.
Their radiative properties
(particularly, multiwavelength spectra) are quite different from those of
active pulsars, but the presense of the thermal
component in their emission
provides a clue to understand the nature of these objects.

First thermally emitting isolated NSs were detected with {\sl Einstein}.
In 1990's, \ros\ and \asc\ detected more than 30 radio pulsars,
including a few thermal emitters,
and discovered several radio-silent NSs
(see Becker and Pavlov\cite{bp02} for a review).
The new era in observing X-ray emission from NSs has
started with the launch of the
{\sl Chandra} and {\sl XMM}-Newton X-ray observatories.
In this brief review we present recent results on
thermal X-ray emission from isolated NSs of
various types observed with \cha.

\begin{figure}[ht]
\centerline{\epsfysize=2.5in\epsfbox{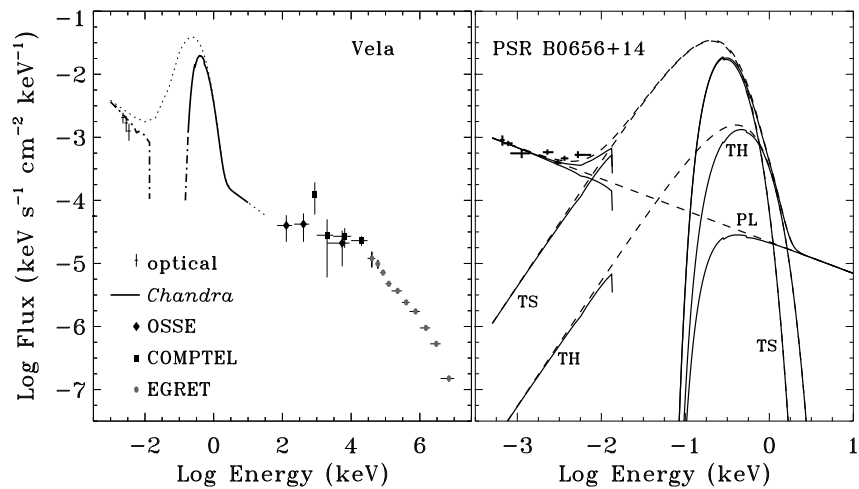} }
\caption{{\it Left:}
Multiwavelength energy spectrum of the Vela pulsar.
The solid line shows the NS hydrogen atmosphere plus PL fit 
to the observed {\sl Chandra} spectrum.
The dotted line is the unabsorbed model flux.
The dash-dot lines show the extrapolated optical and EUV absorbed spectra.
{\it Right:}
Spectral flux for the three-component (TS+TH+PL) model
for PSR B0656+14.  The absorbed and unabsorbed
spectra are shown with solid and dashed curves, respectively.
The crosses show the IR-optical-UV fluxes.
}
\label{vela0656}
\end{figure}

\section{Thermal emission from radio pulsars}
X-ray emission of rotation-powered pulsars generally consists
of two components --- thermal and nonthermal. 
The nonthermal component,
generated by relativistic particles in the NS
magnetosphere, is expected to be strongly pulsed and
characterized by a power-law (PL) spectrum.
The thermal component, emitted from the NS surface,
can be observed in X-rays if it is not
buried under the nonthermal component, as in very young pulsars,
and if the temperature is not too low, as in old pulsars.
Weak pulsations of the thermal emission can be caused by temperature
nonuniformities along the NS surface
and/or anisotropy of emission in a strong magnetic field.
Some examples of thermal radiation of pulsars observed with \cha\
are presented below (see Pavlov et al.\cite{p02a} for more details 
and references).

\cha\ observations of the famous {\em Vela pulsar}
($\tau_c\equiv P/2\dot{P}=11$~kyr, $d\simeq 300$~pc)
made it possible to resolve the pulsar from its surroundings.
The analysis of the \cha\ data revealed
that the pulsar's soft X-ray emission, at $E\lapr 1.8$~keV, is 
dominated by a thermal component (Fig.~\ref{vela0656}, left panel).
The parameters of this component are noticeably different
for the blackbody (BB) fit, $T^\infty_{\rm bb} \simeq 1.4$~MK and
$R^\infty_{\rm bb} \simeq 2.6$~km, and the NS hydrogen atmosphere fit,
$T_{\rm eff}^\infty \simeq 0.7$~MK and $R^\infty \simeq 18$~km
(the superscript $^\infty$ denotes the quantities as measured by
a distant observer).
Such strong difference is explained by the fact that
the hydrogen atmosphere spectra are harder than the BB ones
at the same effective (surface) temperature\cite{zp02}.

The lack of significant spectral lines in the  
spectrum indicates that there are no heavy elements
in the radiating layers.  On the other hand, even strongly
magnetized hydrogen does not have spectral features in the investigated
energy band. The two-component X-ray model involving the NS hydrogen 
atmosphere fits well the optical data on the pulsar (Fig.\ \ref{vela0656}).
In this interpretation, the effective temperature of the Vela pulsar 
is well below the predictions
of the ``standard'' models of NS cooling\cite{t98,y02}.

The analysis of the \cha\ data on {\em PSR B0656+14} 
($\tau_c=110$ kyr) showed no significant lines in the source spectrum.
A three-component model is required to fit the data in the 0.2--6~keV range 
(Fig.\ \ref{vela0656}, right panel).
The high-energy tail ($E\gapr 2$ keV) fits best with a PL spectrum
with a slope close to that of the optical spectrum.
The low-energy part of the spectrum
($E\lapr 0.7$ keV) fits well with a BB model (thermal soft; TS), 
with temperature $T_{\rm s}^\infty \simeq 0.85$~MK and 
radius $R_{\rm s}^\infty\simeq 17$~km (for $d=700$~pc).
This thermal soft (TS) component is presumably emitted from the
whole visible NS surface.  To fit the spectrum at intermediate energies,
an additional component is needed.
A good fit is obtained by adding a thermal hard (TH) component,
with a temperature $T^\infty_{\rm h}\simeq 1.65$ MK
(as given by the BB model), presumably
emitted from pulsar's polar caps of radius
$R_{\rm pc}=R^\infty_{\rm h}\simeq 1.4$~km heated by
relativistic particles streaming down from the pulsar acceleration zones.
The bolometric luminosity of the TS component is
a factor of 10 higher than that for the TH component.
This three-component model fits the data from IR through X-ray energies.

The analysis of the \cha\ data on {\em PSR B1055--52} ($\tau_c=540$~kyr)
yields results very similar to those obtained for PSR B0656+14.
The high-energy tail of the spectrum ($E\gapr 2$ keV) is
best described with a PL, which approximately matches the optical and
gamma-ray data.  The spectrum at lower energies fits well with two thermal (BB)
components, soft and hard.
The best-fit model parameters for the phase-integrated spectrum are
$T^\infty_{\rm s}=0.75$~MK and $R^\infty_{\rm s}=13$~km
for the soft thermal component, $T^\infty_{\rm h}=1.4$~MK
and $R^\infty_{\rm h}= 1.2$ km
for the hard thermal component (for $d=700$~pc).
Interestingly, the temperature
of the soft thermal component is approximately the same for these pulsars,
although the characteristic age, $\tau_c$, is a factor of 5
larger for B1055--52. A possible explanation of this fact is that these
NSs have different properties (e.g., the mass of B1055--52 is lower\cite{y02}).
An alternative explanation\cite{p02a} is that
the true ages of these pulsars can be much closer to each other than
the characteristic ages.

%%%%%%%%%%%%%%% TABLE 1
\begin{table}[ph]
\tbl{Examples of compact central objects in SNRs with best-fit parameters
for the BB  and NS hydrogen atmosphere models
of the thermal component.\vspace*{1pt}}
{\footnotesize
\begin{tabular}{|c|cc|c|cc|cc|}
\hline
 Object & Host SNR & Age & Period &
 $T^\infty_{\rm bb}$ & $R^\infty_{\rm bb}$ 
 &
 $T_{\rm eff}^\infty$ & $R^\infty$ \\
 &  &  kyr & & MK & km 
 &
      MK & km \\
\hline
J2323+5848  & Cas A       & 0.32  & ...     & 5.7 & 0.5     
    &
3.5 & 1 \\
J0852--4617 & G266.1--1.2 & $\sim$1 & ... & 4.6 & 0.3     
    &
3.1 & 1.5 \\
J0821--4300 & Pup A       & 1--3  & ...   & 4.4 & 1.4 
 &
2.0 &  10 \\
J1210--5226 & G296.5+10.0 & 3--20 & 424~ms   & 2.9 & 1.6  
 &
1.6 & 11 \\
\hline
\end{tabular}\label{tab1} }
\vspace*{-13pt}
\end{table}
%%%%%%%%%%%%%%%%%%%%%%%%%%%%%

\section{Compact central objects in supernova remnants}
Among the radio-quiet NSs, there is an interesting subclass dubbed
`compact central objects (CCOs) in supernova remnants'.
Four examples of such objects, studied with \cha, are listed in Table 1.
Although the spectra of CCOs are very similar to each other,
it does not necessarily mean that they represent a uniform class of objects.
If we adopt a plausible hypothesis that the thermal component of
their radiation can be adequately described
by the same spectral model (e.g., BB or hydrogen NS atmosphere),
then we have to conclude that the temperature is decreasing with age
while the emitting area is increasing.
If the radiation emerges from a hydrogen
atmosphere, then the effective radius
is consistent with a NS radius for two oldest
CCOs, J0821--4300 and J1210--5226,
being smaller for two youngest ones, J2323+5848 and J0852--4617.

A recent review on CCOs has been presented
by Pavlov et al.\cite{p02b}. Here we describe briefly only the 
the best-investigated CCO, J1210--5226
(=1E~1207.4--5209) at the center of G296.5+10.0 (=PKS 1209--51/52), 
{\em the first isolated NS which shows spectral lines}.

First \cha\ observation\cite{z00} of J1210--5226 resulted in the
discovery of the period, $P\simeq 424$~ms,
which proved that the source is a NS.
Second \cha\ observation\cite{p02c} and an 
{\sl XMM}-Newton observation\cite{m02} provided an estimate for the
period derivative,  $\dot{P} \approx 3\times 10^{-14}$ s s$^{-1}$.
This estimate implies that the characteristic age
of the NS, $\tau_c \approx 200$ kyr, is much larger than the 3--20 kyr age
of the SNR, while the ``canonical'' magnetic field
(assuming a centered-dipole configuration),
$B\approx 4 \times 10^{12}$  G, is typical for a radio pulsar.

\begin{figure}[ht]
\centerline{\epsfysize=1.6in\epsfbox{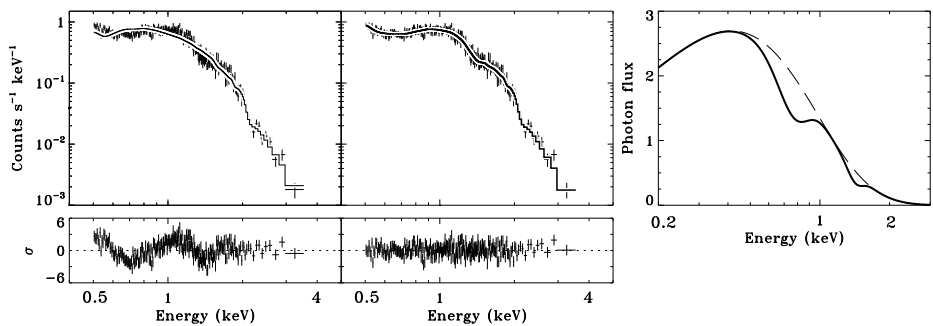}}
\caption{
{\sl Left:} Fit to the \cha\ spectrum of J1210--5226
with a BB
model. {\sl Middle:} The same model with two absorption
lines added (at $E=0.7$ and 1.4~keV).
{\sl Right:} The two-line model for the photon flux
(in $10^{-3}$~s$^{-1}$~cm$^{-2}$~keV$^{-1}$).
}
\label{1207}
\end{figure}

The analysis of the \cha\ spectra\cite{s02}
shows that continuum models (e.g., BB, PL, fully ionized NS atmosphere)
fail to fit the source spectrum, revealing two absorption features near
0.7 keV and 1.4 keV, with equivalent widths of $\sim 0.1$ keV
(Fig.~\ref{1207}).  Most likely, the observed features are  atomic lines
formed in the NS atmosphere. The lines
cannot be explained as emerging from a hydrogen atmosphere
because, at any magnetic field and any reasonable gravitational
redshift, there is
not a pair of strong hydrogen
spectral lines whose energies would match the observed ones.
Sanwal et al.\cite{s02} suggested that these features are the strongest lines
of once-ionized helium in a superstrong magnetic field, (1.4--1.7)$\times
10^{14}$ G. Such a high value can be explained by a strong off-centering
of the magnetic dipole or by the presence of a strong non-dipolar component.
In this interpretation, the observed line energies
correspond to the gravitational redshift  $z=0.12$--0.23, which corresponds to
$R/M=8.8$--14.2 km~$M_\odot^{-1}$.  Hailey \& Mori\cite{hm02} 
suggested an alternative interpretation, that these
line are due to He-like ions of oxygen or neon in a magnetic field somewhat
below $10^{12}$ G. Which (if any) of the two interpretations is correct
will be shown by future observations with high spectral resolution.

%%%%%%%%%%%%%%%%%%%%%%%%%%%%%%%%%%%%%%%%%%%%%%%%%%%%%%%%
\section{``Dim'' isolated neutron stars}
\ros\ discovered seven objects
which are often called ``dim'' or ``truly isolated''
NSs (see Haberl\cite{h03} for a recent review). These sources
are characterized by blackbody-like X-ray spectra
with temperatures around 1~MK.  Low values
of hydrogen column density, $n_{\rm H}\sim 1\times 10^{20}$ cm$^{-2}$,
suggest small distances to these objects.
Their X-ray fluxes,
$f_{\rm x}\sim 10^{-12}$--$10^{-11}$ erg s$^{-1}$ cm$^{-2}$,
correspond to rather low
(bolometric) luminosities, $L\sim 10^{30}-10^{31}$ erg s$^{-1}$
for a fiducial distance of 100 pc. The fluxes
 do not show variations on either short
(minutes-hours) or long (months-years) time scales, and they are much higher
than the optical fluxes,
$f_{\rm x}/f_{\rm opt} > 10^3$.
Such properties strongly suggest that these objects are isolated, nearby
NSs.  Two of them, RX~J0720.4--3125 and J1308.8+2127, show periodic variations,
with surprisingly long periods $P\simeq 8.4$ s and 10.3~s, respectively.
The source powering the thermal radiation of these NSs
has been a matter of debate.
Two main options considered are the internal heat of
a cooling NS and accretion of material from the interstellar medium.
Large proper motions measured for RX~J1856.5--3754\cite{w01}
and RX~J0720.4--3125\cite{m03} (about 330 and
105~mas~yr$^{-1}$, respectively) imply high velocities
of these objects, that makes accretion 
an implausible source of energy for the thermal radiation
from these objects.  Moreover, first estimates of the period derivative
for RX~J0720.4--3125\cite{za02}$^,$\cite{k02} 
also favor models invoking a young (or middle-aged)
cooling NS (a pulsar?) rather than an old accreting NS for this object
because the corresponding strong magnetic field ($B\sim 10^{13}$~G)
is believed to prevent the accretion of material onto the NS surface.

\begin{figure}[ht]
\centerline{\epsfysize=2.5in\epsfbox{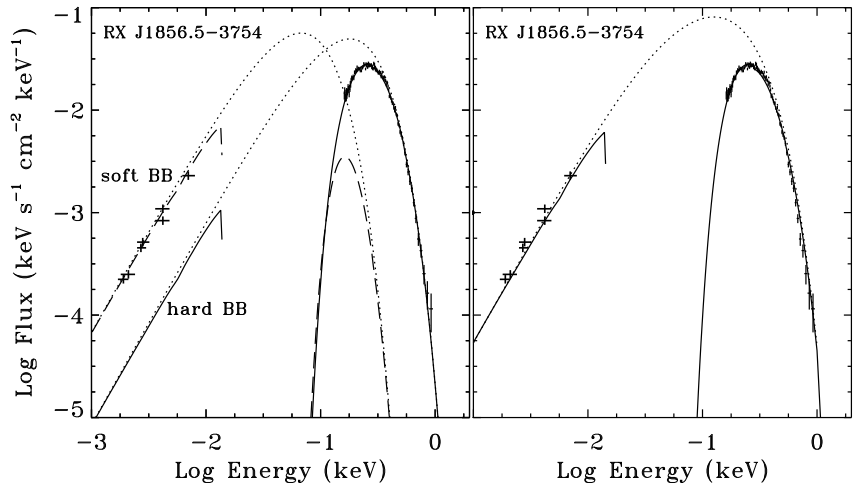}}
\caption{
{\em Left}: Spectral flux of RX~J1856.5--3754 as given by the best BB
fit (``hard BB'' model) to the \cha\ data.
The optical fluxes are fit to a ``soft BB'' model)
(see text for the model parameters).
The dotted lines show the unabsorbed model fluxes.
{\em Right}: Fit of the optical and X-ray data with 
a model assuming a nonuniform (continuous) temperature distribution
over the NS surface (see text). 
}
\label{1856}
\end{figure}

The most famous (and best-investigated)
object of this class, {\it RX~J1856.5--3754} (J1856 hereafter),
discovered by Walter et al.\cite{w96},
has been observed with \cha\ for 505 ks with high spectral resolution.
Detailed analyses of these observations revealed no spectral lines.
The combined X-ray and optical data
rule out all the available NS atmosphere models: light-element models
overpredict the actual optical fluxes\cite{p96},
whereas heavy-element models do not fit the X-ray spectrum.
The spectrum fits well a BB model of $T^\infty_{\rm bb}=0.73$ MK and
$R^\infty_{\rm bb}=4.4$~km for a distance $d=120$~pc, 
as measured from the optical
data (the ``hard BB'' component in Fig.~\ref{1856}, left panel).

Search for pulsations from J1856 at periods
$P>20$~ms has yielded no positive result.
Several attempts have been made to reconcile various heavy-element
atmosphere models with the X-ray data on J1856, assuming fast rotation
($P<10$~ms), which smears spectral features, or very low metal abundances, 
but none of them succeeded.
Thus, we have to admit that the BB spectrum remains the best model
for the X-ray spectrum of J1856.  However, the radius of the emitting area,
$R^\infty_{\rm bb}\simeq 4$--5 km, is too small to be the NS radius.
Moreover, the X-ray BB model 
underpredicts by a factor of 7 the optical fluxes detected from J1856
(Fig.~\ref{1856}).  One may assume\cite{po02} 
that the X-ray radiation is emitted from a hot area on
the NS surface, while the optical radiation is emitted from the
rest of the surface of the NS with a radius $R^\infty_{\rm bb,s}>16$~km and
a temperature, $T^\infty_{\rm bb,s}<0.4$~MK 
(``soft BB'' component in Fig.~\ref{1856}, left panel).
In this two-temperature picture, it is tempting to interpret the hot area
as a pulsar polar cap. This interpretation is supported by 
the discovery\cite{vkk01} of the H$_\alpha$ nebula surrounding J1856,
presumably a bow-shock in the ambient medium
created by the supersonic motion of the NS
with a relativistic (pulsar?) wind.
However, the hot spot is unusually large and luminous
for an ordinary pulsar.  On the other hand, 
a large polar cap would be expected if J1856 were a millisecond
pulsar, with a period $P\sim 2$--5~ms.
Since the surface magnetic filed is low in millisecond pulsars,
it does not prevent the heat released by relativistic particles
to propagate from the ``core'' of the polar cap along the NS surface,
so that the surface temperature decreases gradually away from the core
(see Zavlin and Pavlov\cite{zp98} and Zavlin et al.\cite{z02}). 
To model such a case, we assumed that
a nonuniform temperature distribution over the NS surface can be
approximated as
$T^\infty=T_{\rm hs}^\infty [1+(\theta/\theta_0)^\gamma\,]^{-1}$, where
$\theta$ is magnetic colatitude,
$T_{\rm hs}^\infty$ and $\theta_0$ are characteristic temperature
and angular size of the hot spots, respectively.
We found that such a model, with the spin axis perpendicular to
the line of sight, angle $\alpha=15^\circ$
between the spin and magnetic axes,
$kT_{\rm hs}^\infty=82$~eV, $\theta_0=38^\circ$, $\gamma=2.1$, and
$R^\infty=16.8$~km, satisfactorily fits
the multiwavelength data on J1856 (see Fig.~\ref{1856}, right panel).
In such a model, the nondetection of nonthermal X-ray and radio
emission from J1856 is caused by unfavorable orientation of the
radiating beams, whereas the thermal X-rays are detected from
peripheries of the polar cap(s). To check 
this hypothesis, the source should be observed in X-rays with high
temporal resolution.

Another possible hypothesis\cite{b03} is that the NS surface is composed of a 
solid matter (according to Lai and Salpeter\cite{ls97}, 
this can happen even at high
temperatures if the magnetic field is very strong at the surface,
$B\gapr 10^{14}$ G). The shape of the emission spectrum from the solid
surface could mimic the BB spectrum
in the relatively narrow energy range, 0.1--1 keV, 
but the emissivity would be lower, 
so the radius inferred would be larger, compatible with being a NS
radius. Although there are no reliable calculations for such a model,
we expect that it will give a UV-optical flux even lower than the BB
model, increasing the discrepancy with the observed flux.

\section{Concluding remarks}
Because of the limited space available, we presented only a few examples
of thermal radiation from NSs.  We did not even touch several important
subclasses of thermally emitting NSs, such as ``old'' pulsars,
which apparently emit thermal X-rays from their polar caps\cite{bp02},
radio-quiet ``magnetars'',
which show thermal components\cite{mereg02}, likely due to 
decay of a superstrong magnetic field,
$B\sim 10^{14}-10^{15}$~G, and transiently accreting NSs in quiescence,
whose thermal emission is due to the heat released by pycnonuclear reactions
in the compressed accreting material\cite{rutl02}.
Multiwavelength observations of the diverse population of NS have convincingly
shown that thermal emission indeed dominates
in the soft X-ray and/or UV-optical radiation of various classes of NSs.
The analysis of this emission allows one to measure the NS temperatures
and, in some cases, to estimate the radii or mass-to-radius ratios.
An interesting result of the temperature measurements in middle-aged
pulsars and CCOs, which are supposed to be passively cooling NSs
with relatively well known ages,
is that the age dependence of temperature cannot be described
by a single cooling curve, at least if the current age estimates are not
off too much. It indicates that NSs are not all alike, but at least some
of their properies, responsible for cooling, are different. 
Yakovlev et al.\cite{y02} explain the different cooling behavior as due
to different NS masses. Main cooling regulators in NSs younger than 1 Myr
are the neutrino emission mechanisms (e.g., modified [slow] and direct
[fast] URCA mechanisms at low and high NS core densities, respectively)
and the baryon superfluidity that, generally, reduces the neutrino
emissivity, depending on the poorly known critical temperatures 
$T_{\rm c}(\rho)$ for different types of superfluidity. 
Low-mass NSs (e.g., PSR B1055--52) have lower central densities and cool
slower than high-mass NSs with higher central densities (e.g., Vela).
Moreover, these authors conclude that, for NSs with nucleon cores,
superfluidity is required to reconcile the observations with the cooling
models, although there is not enough data yet to infer 
$T_{\rm c}(\rho)$ and measure the masses of individual NSs. 

The observational data on the NS thermal emission indicate
that the simple picture of a NS with a centered dipole magnetic
field and a uniform surface temperature is, most likely,
an oversimplification. We have seen several examples
of nonuniform temperature distributions, in both active pulsars
and radio-quiet NSs. Moreover, the radius of X-ray emitting area
is often considerably smaller than the ``canonical'' NS radius.
The example of J1210--5226 shows that the characteristic
age of a pulsar can differ from its true age
by a large factor, and the conventional ``pulsar magnetic field''
can be quite different from the actual magnetic field at the NS surface.
Therefore, any inferences on the properties of NSs and the superdense
matter obtained without taking this into account should be considered
with caution.

%%%%%%%%%%%%%%%%%%%%%%%%%%%%%%%%%%%%%%%%%%%%%%%
\section*{Acknowledgments}
We thank our colleagues who participated in the analysis of the above-described
observations, particularly D.\ Sanwal. We are grateful to J.\ Tr\"umper
for the illuminating discussions on the nature of RX~J1856.5--3754
and other ``dim'' neutron stars and D.\ G.\ Yakovlev for his 
enlightening remarks on NS cooling. 
Support for this work was provided by the NASA
through Chandra Awards SP2-2001C, GO2-3088X and GO2-3089X,
issued by the Chandra X-Ray Observatory Center,
which is operated by the Smithsonian Astrophysical Observatory
for and on behalf of NASA under contract NAS8-39073.
This work was also partially supported by NASA grant NAG5-10865.

\end{document}